\documentclass[
	aps,
	twocolumn,
	altaffilletter,
	nolongbibliography,
	numerical,
	flushbottom,
	secnumarabic,
	pra,
	superscriptaddress,
	floatfix,
	nofootinbib,
	10pt
]{revtex4-1}

\usepackage{graphicx}
\usepackage{upgreek}
\usepackage{braket}
\usepackage{amsmath, amssymb}
\usepackage{newtxtext, newtxmath}
\usepackage{siunitx}
\usepackage{booktabs}
\usepackage{subfigure}

\usepackage{makecell}
\usepackage{multirow}
\usepackage{hhline}

\usepackage[breaklinks, pdftex, hyperfootnotes=true, pdfpagelabels, bookmarks, pageanchor]{hyperref}
\hypersetup{%
	colorlinks=true, linktocpage=true, pdfstartpage=1, pdfstartview=FitH, pdfborder={0 0 0},%
	breaklinks=true, pdfpagemode=UseNone, pageanchor=true, pdfpagemode=UseOutlines,%
	plainpages=false, bookmarksnumbered, bookmarksopen=true, bookmarksopenlevel=1,%
	hypertexnames=true, pdfhighlight=/O,
	urlcolor=blue, linkcolor=blue, citecolor=blue, 
}

\usepackage{lipsum}

\newcommand{\ie}{i.\,e.}

\newcommand{\ped}[1]{_\text{#1}}
\newcommand{\api}[1]{^\text{#1}}

\newcommand{\rng}[2]{\ensuremath{[#1, #2]}}
\newcommand{\rngopen}[2]{\ensuremath{(#1, #2)}}
\newcommand{\rnglopen}[2]{\ensuremath{(#1, #2]}}

\newcommand{\nspin}{n}
\newcommand{\n}{N}
\newcommand{\ham}{H}
\newcommand{\hamtf}{V\ped{TF}}

\newcommand{\hamtarget}{\ham_0}
\newcommand{\hamsysbath}{\ham_{SB}}
\newcommand{\tf}{\tau}
\newcommand{\sinv}{s\ped{inv}}
\newcommand{\tinv}{t\ped{inv}}
\newcommand{\scrit}{s\ped{c}}
\newcommand{\tcrit}{t\ped{c}}
\newcommand{\lpause}{t\ped{p}}
\newcommand{\mingap}{\Delta}
\newcommand{\sgap}{s_\mingap}
\newcommand{\tgap}{t_\mingap}
\newcommand{\pgs}{P_0}
\newcommand{\diss}{\mathcal{D}}
\newcommand{\iu}{i}
\newcommand{\eu}{e}
\newcommand{\omegac}{\omega\ped{c}}

\newcommand*{\ev}[1]{\langle #1 \rangle}

\graphicspath{{pictures/}}

\begin{document}

\author{Gianluca Passarelli}
\affiliation{Dipartimento di Fisica ``E.\,Pancini'', Universit\`a degli Studi di Napoli Federico II, Complesso di Monte S.~Angelo, via Cinthia - 80126 - Napoli, Italy}
\affiliation{CNR-SPIN, c/o Complesso di Monte S. Angelo, via Cinthia - 80126 - Napoli, Italy}

\author{Ka-Wa Yip}
\affiliation{Department of Physics, University of Southern California, Los Angeles, CA 90089}
\affiliation{Center for Quantum Information Science \& Technology, University of Southern California, Los Angeles, CA 90089}

\author{Daniel A. Lidar}
\affiliation{Center for Quantum Information Science \& Technology, University of Southern California, Los Angeles, CA 90089}
\affiliation{Departments of Electrical and Computer Engineering, Chemistry, and Physics, University of Southern California, Los Angeles, CA 90089}

\author{Hidetoshi Nishimori}
\affiliation{Institute of Innovative Research, Tokyo Institute of Technology, Yokohama, Kanagawa 226-8503, Japan}
\affiliation{Graduate School of Information Sciences, Tohoku University, Sendai, Miyagi 980-8579, Japan}
\affiliation{RIKEN Interdisciplinary Theoretical and Mathematical Sciences (iTHEMS), Wako, Saitama 351-0198, Japan}

\author{Procolo Lucignano}
\affiliation{Dipartimento di Fisica ``E.\,Pancini'', Universit\`a degli Studi di Napoli Federico II, Complesso di Monte S.~Angelo, via Cinthia - 80126 - Napoli, Italy}

\title{Reverse quantum annealing of the $p$-spin model with relaxation}
\begin{abstract}
In reverse quantum annealing, the initial state is an eigenstate of the final problem Hamiltonian and the transverse field is cycled rather than strictly decreased as in standard (forward) quantum annealing. We present a numerical study of the reverse quantum annealing protocol applied to the $p$-spin model ($p=3$), including pausing, in an open system setting accounting for dephasing in the energy eigenbasis, which results in thermal relaxation. We consider both independent and collective dephasing and demonstrate that in both cases the open system dynamics substantially enhances the performance of reverse annealing. Namely, including dephasing overcomes the failure of purely closed system reverse annealing to converge to the ground state of the $p$-spin model. We demonstrate that pausing further improves the success probability. The collective dephasing model leads to somewhat better performance than independent dephasing. The protocol we consider corresponds closely to the one implemented in the current generation of commercial quantum annealers, and our results help to explain why recent experiments demonstrated enhanced success probabilities under reverse annealing and pausing.
\end{abstract}

\maketitle

\section{Introduction}\label{sec:intro}

Many binary combinational optimization tasks, including the traveling salesman problem, number and graph partitioning, Boolean satisfiability, prime factorization, search tasks, and many others~\cite{cook:optimization}, can be rephrased as finding the ground state of an Ising spin system~\cite{lucas:np-complete}. The resulting spin glass Hamiltonians may involve long-range and/or $p$-body interactions with $p \geq 2 $, as for satisfiability problems~\cite{mezard-montanari}; a reduction to $p=2$ is always possible, but comes at the expense of using ancilla spins~\cite{Cao:2014}. Finding the ground state of these Hamiltonians is NP-hard, colloquially due to the presence of many local minima in the cost function.
In general, efficient algorithms for solving this class of problems are not known, or are in practice beyond the computational power of high-performance computers. Depending on the problem, there may exist heuristic methods that are usually able to produce approximate solutions of the optimization task. Among these, there are greedy algorithms and local searches~\cite{cormen:algorithms}, evolutionary algorithms~\cite{goldberg:genetic-algorithms, passarelli:genetic}, simulated annealing~\cite{kirkpatrick:sa} and its quantum version, quantum annealing~\cite{farhi:quantum-computation, albash:review-aqc, kadowaki:qa, dickson:thermal-qa}.

These heuristic methods offer no control, \textit{a priori}, on the accuracy of their sub-optimal output. One way to improve their efficiency is to use them as steps of a multistage optimization process, where the output of a stage is used as input of the subsequent one. This iterative process usually produces more refined solutions. In this context, reverse quantum annealing has been proposed and studied as a viable tool for multistage optimization~\cite{perdomo:sombrero,chancellor:reverse,marshall,nishimori:reverse-pspin} and quantum simulation~\cite{King:2018aa,King:2019aa}, though its origins can be traced to the very first quantum annealing experiment~\cite{Brooke1999}.

Reverse annealing is a relatively novel global-control feature of the D-Wave quantum annealers~\cite{dwave-site}. In conventional quantum annealing, the system starts in a uniform superposition of computational basis states, and evolves subject to monotonically decreasing quantum fluctuations to target the wanted solution. In contrast, in reverse annealing the system is prepared in a state supposedly close to the correct solution. For instance, this state can be the output of another optimization routine. Quantum fluctuations are first increased, up to an inversion point during the dynamics, and then decreased. If the inversion point is chosen well, the output is an improved trial solution, \ie, a quantum state having larger overlap with the correct one. Reverse annealing can also be combined with the pausing features of the D-Wave machines, allowing to stop the annealing for an extended time period to favor relaxation towards the ground state~\cite{marshall}. 

Reverse quantum annealing and pauses can lead to a significant enhancement of the success probability of quantum annealing compared with the usual protocol, as demonstrated by some recent theoretical and experimental papers concerning these two strategies~\cite{nishimori:reverse-pspin, nishimori:reverse-pspin-2, passarelli:pausing, marshall,King:2018aa,King:2019aa}. However, experimental works are limited to low-connectivity Ising systems due to current hardware limitations. Strongly connected models, which often encode interesting optimization problems~\cite{lucas:np-complete,Venturelli:2014nx}, do not fit natively in the Chimera graph of D-Wave machines, and require minor embedding~\cite{choi:2008} or other reduction methods~\cite{zoller:many-body-into-pairs, dodds:practical-designs, passarelli:genetic}, incurring in all cases a significant overhead in terms of ancillary degrees of freedom. Numerical simulations avoid this overhead and can help to shed light on the behavior of these systems.

In this work, we apply reverse annealing and pauses to the fully-connected ferromagnetic $ p $-spin model~\cite{derrida:p-spin, gross:p-spin, bapst:p-spin}. In the thermodynamic limit, this model encodes a Grover-like adiabatic search~\cite{grover:search, roland-cerf, RPL:10} for odd $ p \to \infty $ ($ p \leq \nspin $, where $\nspin$ is the number of qubits; see Appendix B of Ref.~\cite{passarelli:pausing}). Despite being exactly solvable, the $ p $-spin model has a non-trivial phase diagram deeply related to NP-hard optimization. In particular, this model is subject to a first-order quantum phase transition in the thermodynamic limit for $ p > 2 $~\cite{bapst:quantum-spin-glass}. At the critical point, the gap $ \mingap $ between the ground state and the first excited state closes exponentially in the system size $\nspin$. The annealing time has to be large on the time scale dictated by $ \mingap^{-1} $. Hence, QA is expected to be highly inefficient in finding the ground state of large instances of this model for $p>2$~\cite{wauters:pspin}.

In Ref.~\cite{nishimori:reverse-pspin}, the static properties of the $ p $-spin model were studied using mean field theory.
In particular, the authors focused on the protocol they called adiabatic reverse annealing (ARA), whereby 
an additional parameter $ \lambda $ determines the strength of the starting Hamiltonian and that of the transverse field. For initial states sufficiently close to the ferromagnetic state, there exist paths  in the phase diagram that avoid first-order quantum phase transitions, thus providing an exponential speed-up compared with the standard, forward annealing case. The 
dynamics of the ARA protocol was subsequently studied in Ref.~\cite{nishimori:reverse-pspin-2}
by numerically solving the Schr\"odinger equation and the conclusions based on the static analysis were confirmed. The ARA protocol as studied in Refs.~\cite{nishimori:reverse-pspin,nishimori:reverse-pspin-2} has not yet been implemented in physical quantum annealers.

A second protocol studied in Ref.~\cite{nishimori:reverse-pspin-2} is
iterated reverse annealing (IRA), a reverse annealing protocol that is very similar to the one implemented in the current generation of D-Wave devices, and which we focus on in the present work. Ref.~\cite{nishimori:reverse-pspin-2} studied IRA in the setting of a closed system undergoing unitary evolution, and found that this protocol fails to improve the solution of the ferromagnetic $ p $-spin model with $ p = 3 $, at least for the particular annealing schedule  adopted there. Here, we test a more experimentally realistic annealing schedule and confirm that  in this case too, there are no significant advantages in using IRA, compared to standard quantum annealing, in the unitary limit. 	
However, 
we show that relaxation mechanisms associated with open system dynamics can strongly modify the final outcome. They indeed help in reaching the desired ground state, thus improving the efficiency of IRA. This helps to explain why experimentally the IRA protocol has been observed to be beneficial~\cite{marshall}.

This paper is organized as follows. In Section~\ref{sec:reverse}, we describe our reverse annealing protocol. In Section~\ref{sec:pspin}, we present the $ p $-spin Hamiltonian. For $ p > 3 $, nonstoquastic catalysts are known to turn first-order quantum phase transitions (QPTs) into second-order ones~\cite{seoane:transverse-interactions}, where the gap closes polynomially as a function of $ \nspin $. This improves the scaling of the time-to-solution for this model. In this sense, $ p = 3 $ is the hardest case for quantum annealing.
In Section~\ref{sec:unitary}, we show the results for unitary reverse annealing. Here, we study a system of $ \nspin = 20 $ qubits with $ p = 3 $. We show that the probability of ending up in the correct ground state, i.e., the success probability, depends on the inversion point, and on the magnetization of the initial state. This is in agreement with previous findings on this model~\cite{nishimori:reverse-pspin, nishimori:reverse-pspin-2}. 

Realistic quantum processors are open systems coupled to their environment and are subject to decoherence~\cite{mishra:finite-temperature-qa}. In Section~\ref{sec:dissipative}, we study the dephasing dynamics of this system within the Born-Markov approximation, using a time-dependent Monte Carlo wavefunction approach~\cite{yip:mcwf, passarelli:pausing}, and we compare two different models of dephasing: independent and collective. When the qubits are coupled to a single, collective bath, we show that the success probability of reverse annealing does not depend on the choice of the initial magnetization. In Sec.~\ref{sec:pausing}, we also address the effect of pausing at the inversion point, and show that pauses can improve performance for both models of dephasing.
We present our conclusions in Section~\ref{sec:conclusions}.

\section{Reverse annealing}
\label{sec:reverse}
	
\begin{figure*}[t]
	\subfigure[]{\includegraphics[width = \columnwidth]{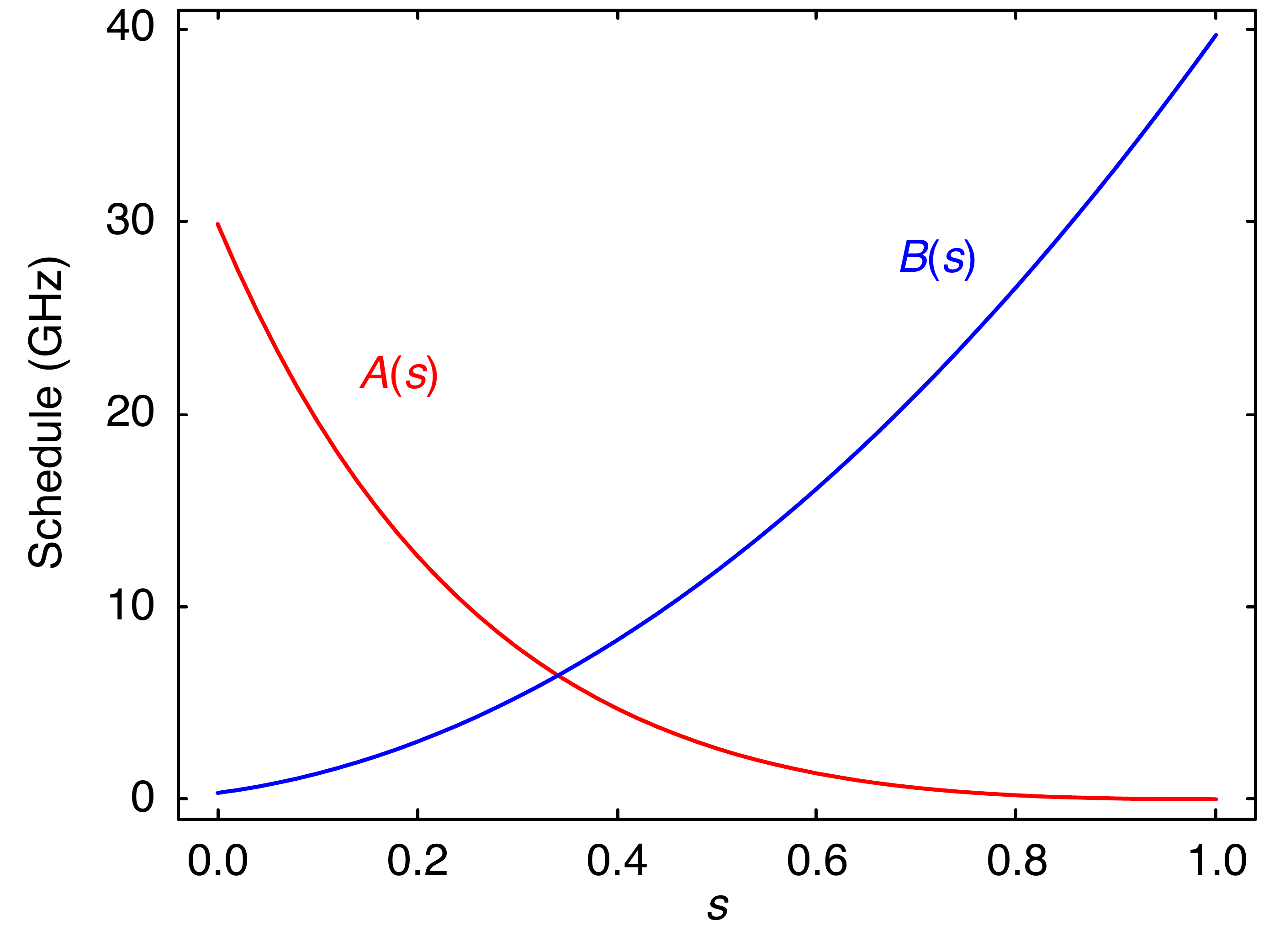}\label{fig:schedules}}
	\subfigure[]{\includegraphics[width = \columnwidth]{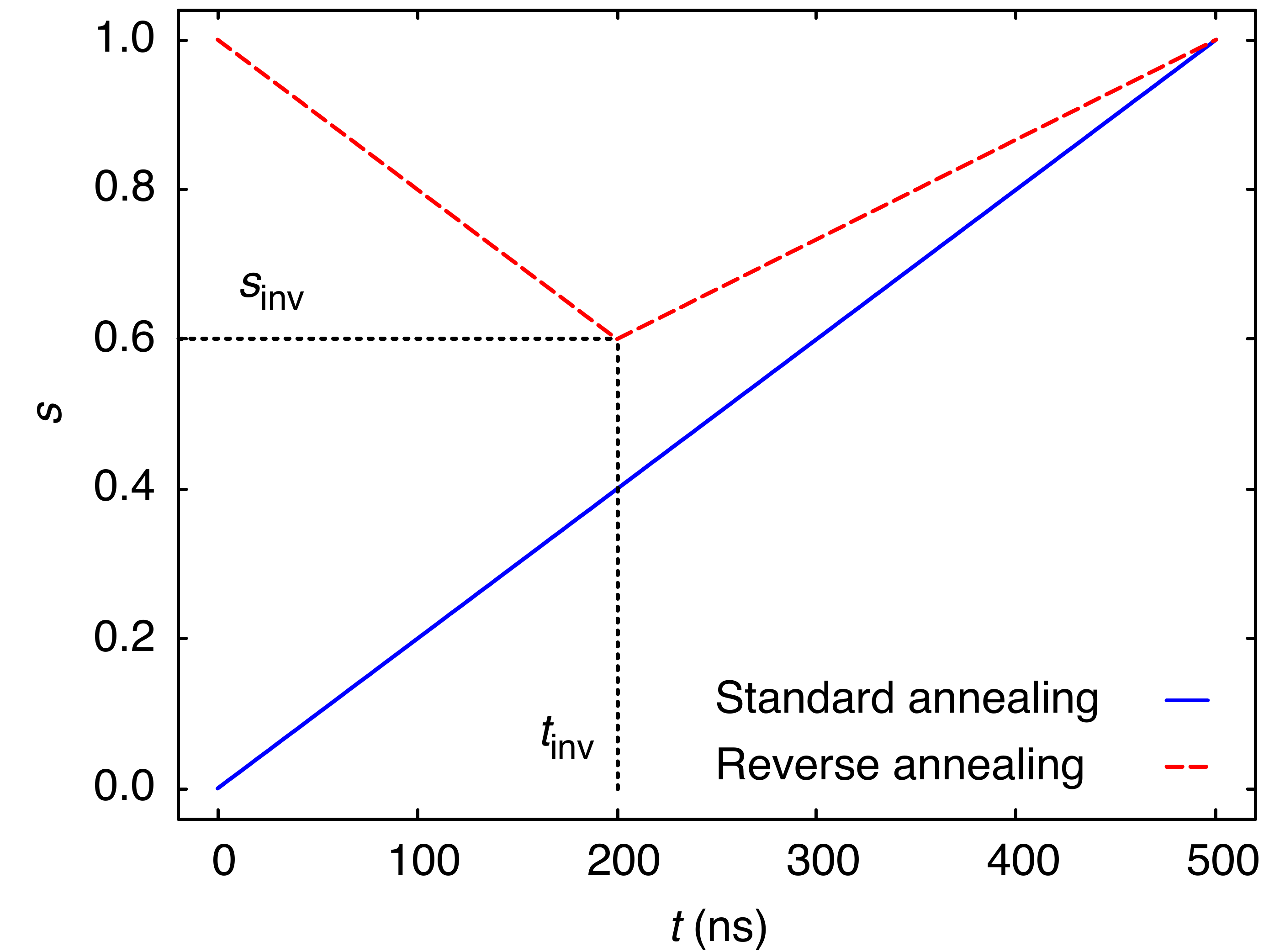}\label{fig:s-of-t}}
	\caption{(a) Annealing schedules (in units such that $\hbar=1$) as a function of the annealing fraction $s(t)$, chosen to be similar to the schedules of the D-Wave processors.
	(b) Annealing fraction $ s(t) $. The blue solid curve represents standard forward quantum annealing of total annealing time $ \tf = \SI{500}{\nano\second} $. The red dashed curve represents reverse annealing of the same total duration with an inversion point $ (\tinv = \SI{200}{\nano\second}, \sinv = 0.6) $.}
\end{figure*}	
	
Standard quantum annealing aims at solving optimization problems by employing quantum fluctuations that are slowly decreased to zero, to efficiently explore the solution space~\cite{kadowaki:qa, santoro-martonak, albash:review-aqc}. A system of $ \nspin $ qubits is prepared at $ t = 0 $ in the ground state of a transverse field Hamiltonian, \ie, the uniform superposition over the $ \n = 2^\nspin $ computational basis states $ \{\ket{0}, \ket{1}, \dots, \ket{\n - 1}\} $. 
The magnitude of transverse field
is then slowly decreased, while the magnitude of the Hamiltonian $ \hamtarget $, encoding the optimization problem, is simultaneously increased. The adiabatic theorem guarantees that if the evolution is long on the timescale set by the inverse of the minimal gap $ \mingap = \min_t [E_1(t) - E_0(t)] $ between the ground state and the first excited state (we set $ \hslash = 1 $ henceforth), then at the end of the anneal $ t = \tf $ the system populates the target ground state of $ \hamtarget $ with a probability $ \pgs $ that approaches unity~\cite{Jansen:07,lidar:102106}. However, any finite sweep rate leads to diabatic Landau-Zener transitions at the avoided crossings~\cite{Joye:LZ}, thus reducing the success probability $ \pgs $ of the adiabatic algorithm. Therefore, the output state of quantum annealing is in general a trial solution of the optimization problem, ideally having a large overlap with the target ground state.
	
Reverse annealing instead aims at refining an already available trial solution~\cite{perdomo:sombrero}. For instance, NP-hard optimization tasks are solved using heuristics, whose output is often an approximation of the true global minimum. The algorithm of reverse annealing is the following.
\begin{enumerate}
	\item At $ t = 0 $, the system is prepared in the trial solution state.
	\item Quantum fluctuations are increased, until a turning point $ \tinv $ is reached during the dynamics. This ends the reversed part of the dynamics.
	\item After the turning point, the dynamics follows the standard quantum annealing schedule; quantum fluctuations are decreased until $ t = \tf $, when the state is eventually measured.
\end{enumerate}
Careful choices of the turning point, and of the initial state, can lead to an improvement in the solution. Moreover, this scheme can also be repeated multiple times, each time starting from the output of the previous stage; hence the terminology of iterated reverse annealing. Assuming that each iteration improves the success probability $ \pgs $, this procedure can systematically improve the quality of the solution. 
	
We consider the following time-dependent Hamiltonian, suitable for IRA but not ARA (which includes another term):
\begin{equation}
\label{eq:time-dependent-hamiltonian}
	\ham(t) = A\bigl(s(t)\bigr) \hamtf + B\bigl(s(t)\bigr) \hamtarget, \quad  \hamtf = -\sum_{i=1}^{\nspin} \sigma_i^x .
\end{equation}
Since we only consider the IRA protocol, henceforth we simply refer to it as reverse annealing. The function of time $ s(t) $ in Eq.~\eqref{eq:time-dependent-hamiltonian} is the annealing fraction (or dimensionless time) and satisfies $ 0 \le s(t) \le 1 $ for all $ t $. The two functions $ A(s) $ and $ B(s) $ determine the annealing schedule, which we choose to match the annealing schedule of the D-Wave processors [see Fig.~\ref{fig:schedules}]. They satisfy $ A(0) \gg B(0) $ and $ B(1) \gg A(1) $.
	
The functional form of $ s(t) $ distinguishes between standard (forward) and reverse annealing. In standard quantum annealing the dimensionless time is defined as $ s(t) = t / \tf $, $ \tf $ being the annealing time. Thus, $ s(t) $ is a monotonic function of $ t $, and is represented in the plane $ (t, s) $ by a straight line going from $ (0, s_0) $ to $ (\tf, s_1) $, where $ s_0 = s(0) = 0 $ and $ s_1 = s(\tf) = 1 $. This is shown in Fig.~\ref{fig:s-of-t} using a blue solid line, for $ \tf = \SI{500}{\nano\second} $. During standard quantum annealing, quantum fluctuations are very large at $ t = 0 $, and decrease monotonically until $ t = \tf $.
	
In contrast, in reverse annealing $ s_0 = s_1 = 1 $. Starting from $ s = s_0 $, where quantum fluctuations are zero, the annealing fraction is first decreased until it reaches the inversion point, $ s = \sinv $, at a time $ t = \tinv $. In this first branch, quantum fluctuations are increased. At $ s = \sinv $, the annealing fraction is then increased towards $ s = s_1 $, and quantum fluctuations are decreased again to zero. In Fig.~\ref{fig:s-of-t}, we show a typical function $ s(t) $ for a reverse annealing of annealing time $ \tf = \SI{500}{\nano\second} $ with an inversion point $ (\tinv = \SI{200}{\nano\second}, \sinv = 0.6) $, using a red dashed line.
		
In general, $ \sinv $ and $ \tinv $ can be chosen independently of each other. However, in this work we choose the following linear relation, in order to have only one free parameter:
\begin{equation}
\label{eq:inversion-time}
	\tinv = \tf (1 - \sinv)\ , \ \ \sinv \ne 0, 1.
\end{equation}
In this way, we have that
\begin{equation}\label{eq:schedule}
	s(t) =
	\begin{cases}
		1-t / \tf & \text{for $ t \le \tinv $},\\
		\frac{1 - \sinv}{\tf \sinv} t + \frac{2 \sinv - 1}{\sinv} & \text{for $ t > \tinv $}.
	\end{cases}
\end{equation}
	
Another possible choice would be to fix $ \tinv $ so that the two slopes are the same, \ie, $ \tinv = \tf / 2 $ for all choices of $ \sinv $. This is similar in spirit to what is discussed in Ref.~\cite{nishimori:reverse-pspin-2}. 
	
In what follows, we will focus on the fully-connected ferromagnetic $ p $-spin model, a model with a permutationally invariant Hamiltonian and a nontrivial phase diagram, often used as a benchmark for the performance of quantum annealing~\cite{gross:p-spin, derrida:p-spin}. 
	
\section{Ferromagnetic $ p $-spin model}\label{sec:pspin}

The Hamiltonian of the ferromagnetic $ p $-spin model, in dimensionless units, reads
\begin{equation}\label{eq:pspin}
	\hamtarget = -\frac{\nspin}{2}{\left(\frac{1}{\nspin} \sum_{i = 1}^{\nspin} \sigma^z_i\right)}^p,
\end{equation}
with $ p \ge 2 $. For even $ p $, there are two degenerate ferromagnetic ground states, whereas for odd $ p $ the ferromagnetic ground state is nondegenerate. For $ p = 2 $, the Hamiltonian of Eq.~\eqref{eq:time-dependent-hamiltonian} is subject to a second-order QPT in the thermodynamic limit at the critical point $ \scrit $ (or, equivalently, $ \tcrit = \tf \scrit $), separating a para- and a ferromagnetic phase. For $ p > 2 $, the QPT is first-order~\cite{bapst:p-spin}. The presence of QPTs affects also the finite-size behavior of the system, as the minimal spectral gap $ \mingap $, found at $ \sgap $ ($ \tgap = \tf \sgap $), closes as $ \nspin^{-1/3} $ for $ p = 2 $ or exponentially in $ \nspin $ for $ p > 2 $. $ \sgap $ (\ie, $ \tgap $) approaches $ \scrit $ (\ie, $ \tcrit $) as $ \nspin \to \infty $. First-order QPTs are especially detrimental for quantum annealing, as the annealing time has to grow exponentially with the system size to compensate the closure of the gap at $ s = \sgap $. In the following, we will focus on the case $ p = 3 $.
	
We can define the total spin operator $\mathbf{S} = (S_x,S_y,S_z)$ and the dimensionless magnetization operators
\begin{equation}
	m_{\alpha} = \frac{1}{\nspin} S_{\alpha} ,\quad S_{\alpha} = \sum_{i = 1}^{\nspin} \sigma_i^{\alpha}, \quad \alpha\in\{x,y,z\}
\end{equation}
that allow the rewriting of the time-dependent Hamiltonian of Eq.~\eqref{eq:time-dependent-hamiltonian} as
\begin{equation}
	 \ham(t) = -\frac{A(t)}{2} S_x - \frac{B(t)n}{2} m_z^p .
\end{equation}
Since $[\mathbf{S}^2,S_z]=0$ and both are Hermitian, they share an orthonormal basis $\{\ket{S,\mu_S}\}$ such that the eigenvalues of $\mathbf{S}^2$ are $S(S+1)$ with $S\in\{0,1/2,1,\dots,\nspin/2\}$ for even $ \nspin $ and $S\in\{1/2,1,\dots,\nspin/2\}$ for odd $ \nspin $, and the eigenvalues of $S_z$ are $\mu_S \in \{-S,-S+1,\dots,S\}$.
In the subspace with maximum spin $S=\nspin/2$, we instead label the basis states as $ \ket{w} \equiv \ket{\nspin/2 - w} $, with $ w \in \{0, 1, \dots, \nspin\} $. These are the eigenstates of $ m_z $ with eigenvalues $ m = 1 - 2 w / \nspin $. The target state is the ferromagnetic ground state $ \ket{0} $, \ie, the eigenstate of $ m_z $ with eigenvalue $ m = 1 $. 

The Hamiltonian of the $ p $-spin model commutes with $\mathbf{S}^2$. Hence sectors differing by $S$ do not become coupled under the dynamics generated by the $ p $-spin model Hamiltonian. Since the ferromagnetic ground state and the initial one belong to the subspace with maximum spin $ S = \nspin / 2 $, the interesting dynamics occurs in this subspace, whose dimension scales \emph{linearly} with the number of qubits: $ \n = \nspin + 1 $. This fact enables us to perform numerical calculations with relative large numbers of qubits $\nspin$.

In the following, we will start reverse annealing in each of the $ \nspin $ excited states $ \{\ket{1}, \dots, \ket{\nspin} \}$ of $ \hamtarget $ in the symmetric sector with $ S = \nspin/2 $. The similarity to the ferromagnetic ground state is quantified by the corresponding starting eigenvalue of $ m_z $, denoted $ m_0 $. Note that the $ w $th excited state differs from the ferromagnetic ground state by $ w $ spin flips. Therefore, the initial state and the target solution differ by a fraction $ c = \nspin_\uparrow / \nspin = 1 - w /\nspin $ of up-aligned qubits. These parameters are also related to the Hamming distance $ d\ped{H} $, 
via $ d\ped{H} = \nspin - \nspin_\uparrow = \nspin (1 - c) $.

\section{Unitary dynamics}
\label{sec:unitary}

\begin{figure*}[tb]
	\subfigure[]{\includegraphics[width = \columnwidth]{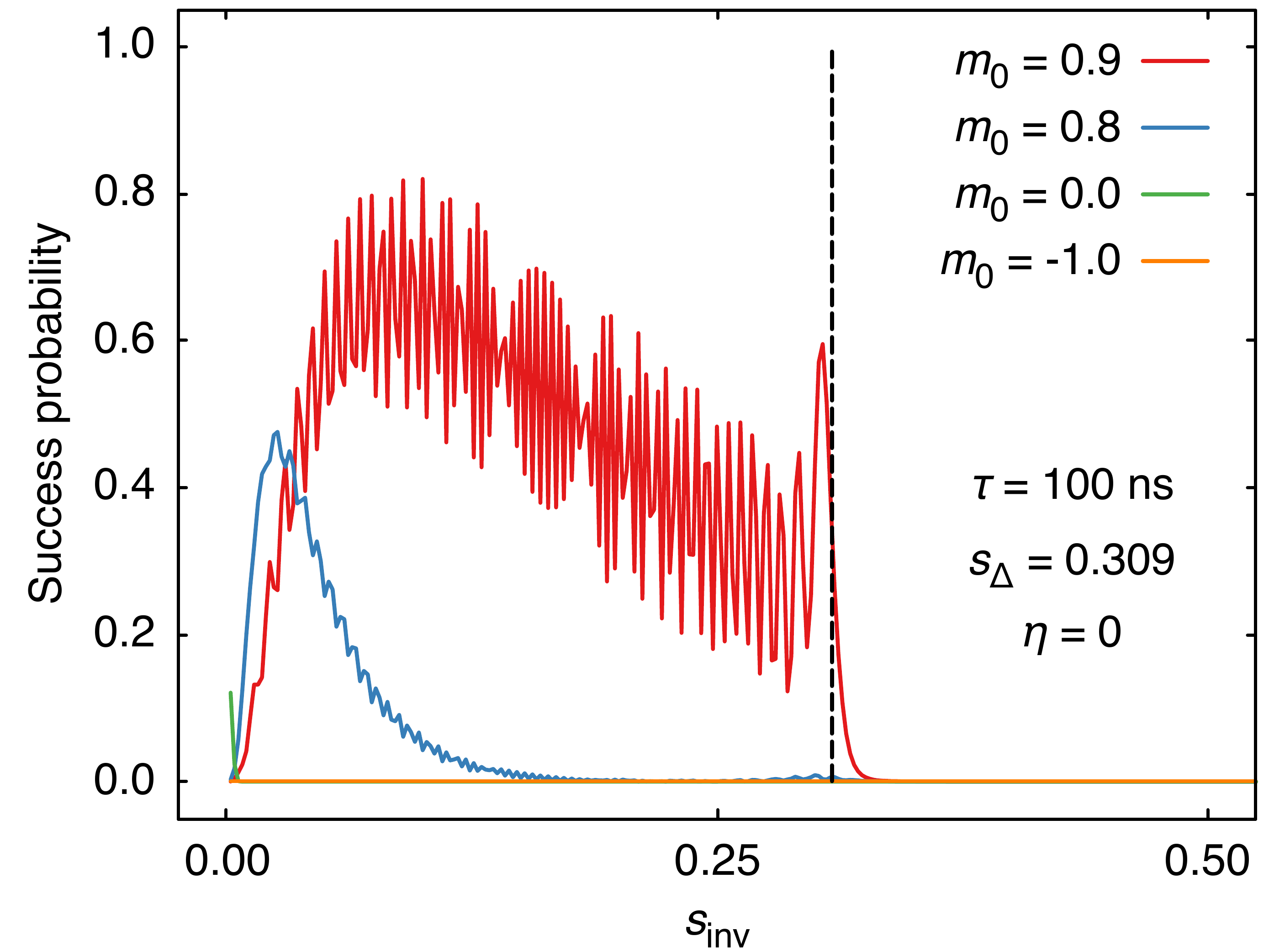}\label{fig:unitary-tf-100}}
	\subfigure[]{\includegraphics[width = \columnwidth]{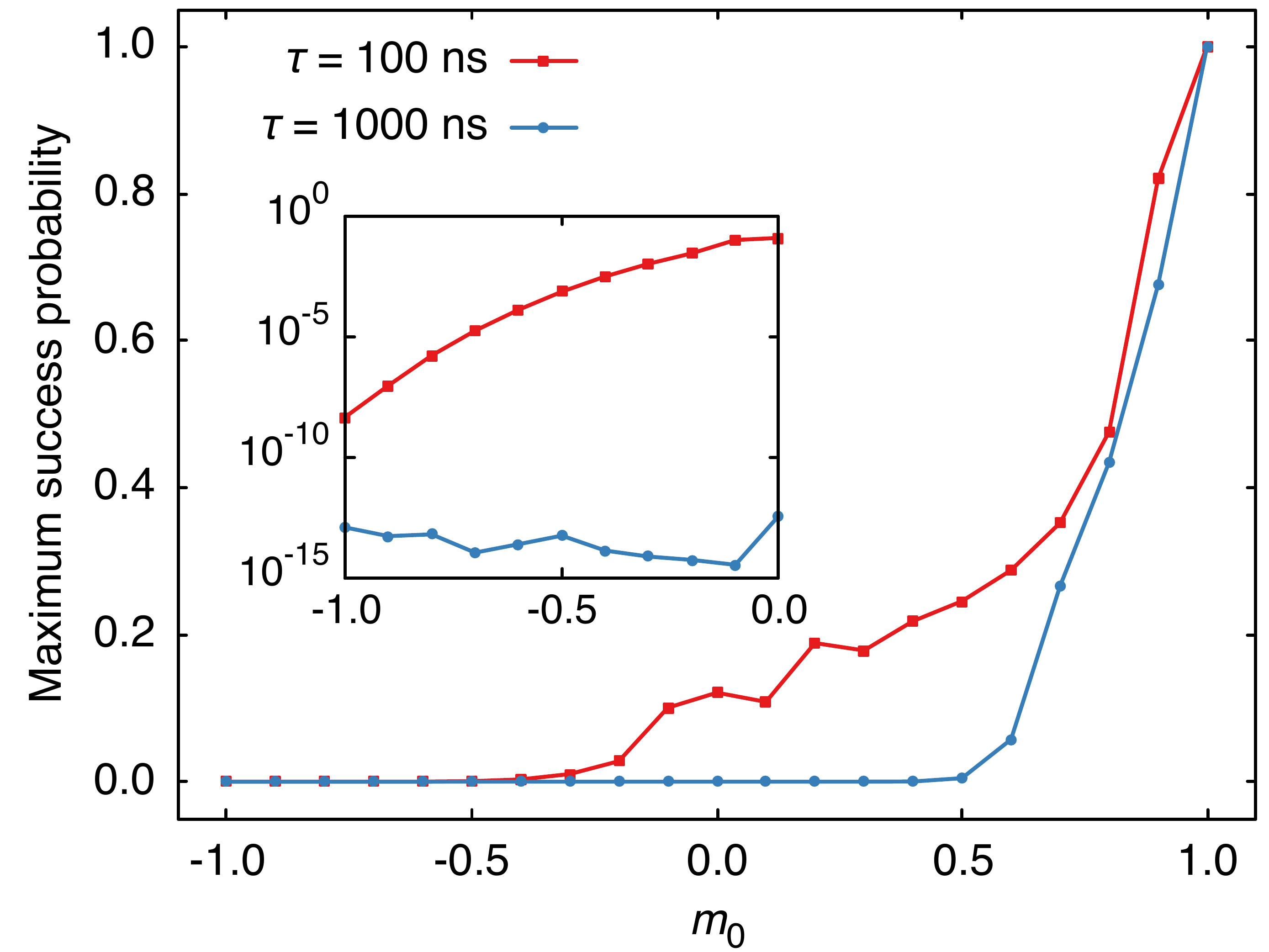}\label{fig:unitary-scaling}}
	\caption{(a) Success probability in unitary reverse annealing as a function of the inversion point $ \sinv $, for several values of the magnetization of the initial state. The dashed vertical line indicates $ \sinv = \sgap \approx 0.309 $. The annealing time is $ \tf = \SI{100}{\nano\second} $. We sampled the interval $ \sinv \in \rngopen{0}{1} $ using a step size of $ \Delta s = 0.002 $, and repeated the dynamics for each choice of $ \sinv $. (b) Maximum success probability achievable with unitary reverse annealing, as a function of the magnetization of the initial state, for two annealing times: $ \tf = \SI{100}{\nano\second} $ and $ \tf = \SI{1000}{\nano\second} $. The inset zooms in on the region $ m\in\rng{-1.0}{0.0} $. Note the logarithmic scale on the vertical axis.}
\end{figure*}

In this section we study the closed system case of a system of $ \nspin = 20 $ qubits, with $ p = 3 $. For our choice of parameters and in terms of the annealing schedules shown in Fig.~\ref{fig:schedules}, the $ p $-spin system has a minimal gap $ \mingap \approx \SI{2.45}{\giga\hertz} $ at $ \sgap \approx 0.309 $. The annealing time is $ \tf = \SI{100}{\nano\second} $. 
	
In Fig.~\ref{fig:unitary-tf-100}, we report the ground state population $ \pgs $  at $ t = \tf $, as a function of the inversion point $ \sinv $, for several initial states: $ m_0 = \text{\numlist{0.9;0.8;0;-1}} $. Recall that the target ground state $ \ket{0} $ has $m=1$. We focus on the region $ \sinv \in \rnglopen{0.0}{0.5}$.
	
The rightmost part of Fig.~\ref{fig:unitary-tf-100} corresponds to cases in which the anneal is reversed too early, \ie, for $ \tinv < \tgap $ and $ \sinv > \sgap $. The system does not cross its quantum critical point, and the success probability is zero. Therefore, no effects on the outcome of the procedure are visible, as the dynamics is slow compared with the minimal inverse level spacing and diabatic transitions are exponentially suppressed. Thus, the system is forced to stay in its initial state, or transition to other excited states. In fact, avoided crossings between pairs of excited eigenstates occur at $ s > \sgap $ for this model, and Landau-Zener processes can further excite the $ p $-spin system.
	
On the other hand, if $ \sinv < \sgap $ the system crosses the minimal gap twice. Here, the success probability benefits from Landau-Zener processes, inducing transitions towards the ground state. In this region, we also note some non-adiabatic oscillations of the success probability, due to the finite annealing time. These oscillations are more evident for large $ m_0 $. As expected from the adiabatic theorem, they are suppressed for longer annealing times. For instance, we verified that they are no longer visible for $ \tf = \SI{1000}{\nano\second} $ (not shown). The sharp rise of the success probability for $ m_0 = 0.9 $ occurs exactly at $ \sinv = \sgap $. For smaller values of $ m_0 $, the success probability rises more smoothly, as the ground state is reached after a preliminary sequence of Landau-Zener transitions between pairs of excited states, whose corresponding avoided crossings occur at $ s > \sgap $. For $ m_0 = 0.8 $, a very small rise of the success probability can still be observed around $ \sinv = \sgap $. This is due to the fact that during the reverse annealing, the system prepared in the second excited state first encounters an avoided crossing with the first excited state, where part of the population is transferred to the latter, and then the avoided crossing with the ground state, where the system populates its ground state. After reversing the dynamics, the two avoided crossings are encountered again (in the reverse order) and part of the population gets excited, thus reducing the success probability $ \pgs $.

As expected, reverse annealing is more effective when the initial state is close to the correct ground state. Moreover, as is also clear from Fig.~\ref{fig:unitary-tf-100}, the inversion time $ \sinv $ must be increasingly close to $0$ for decreasing $ m_0 $, in order to obtain a nonzero success probability at $ t = \tf $. This means that almost the entire dynamics is spent in the reverse part of the annealing, and the system is eventually quenched towards $ s = s_1 $ for $ t \approx \tf $. Even so, if the initial state is too far in energy from the correct solution, the success probability of reverse annealing is always close to zero, as evident from the curves for $ m_0 = 0 $ and $ m_0 = -1 $ in Fig.~\ref{fig:unitary-tf-100}. 
	
The maximum success probability decreases rapidly as a function of $ m_0 $. This is clearly seen in Fig.~\ref{fig:unitary-scaling}, where we report the maximum attainable success probability as a function of $ m_0 $, for annealing times $ \tf = \text{\SIlist{100;1000}{\nano\second}} $. 
Increasing the annealing time reduces non-adiabaticity and results in a lower success probability, compared with that at the end of a faster reverse anneal. As shown in Fig.~\ref{fig:unitary-scaling}, which zooms in on the region $ m_0 \in \rng{-1}{0} $, this decrease can be of several orders of magnitude for poorly chosen trial solutions.  The influence of the annealing time is less pronounced close to $ m_0 = 1 $, and more evident for intermediate and lower values of $ m_0 $. This is consistent with the adiabatic theorem, since a longer anneal time guarantees that the system will have a higher probability of remaining close to the initial eigenstate it has the largest overlap with (not necessarily the ground state)~\cite{Jansen:07}.
	
The results of this Section are in agreement with those reported in Ref.~\cite{nishimori:reverse-pspin-2}. Namely, as is clear from Fig.~\ref{fig:unitary-scaling}, upon iteration the IRA protocol will only decrease the success probability under unitary, closed system dynamics, unless the initial state was already chosen as the solution of the optimization problem. 
	
\begin{figure*}[t]
	\subfigure[]{\includegraphics[width = \columnwidth]{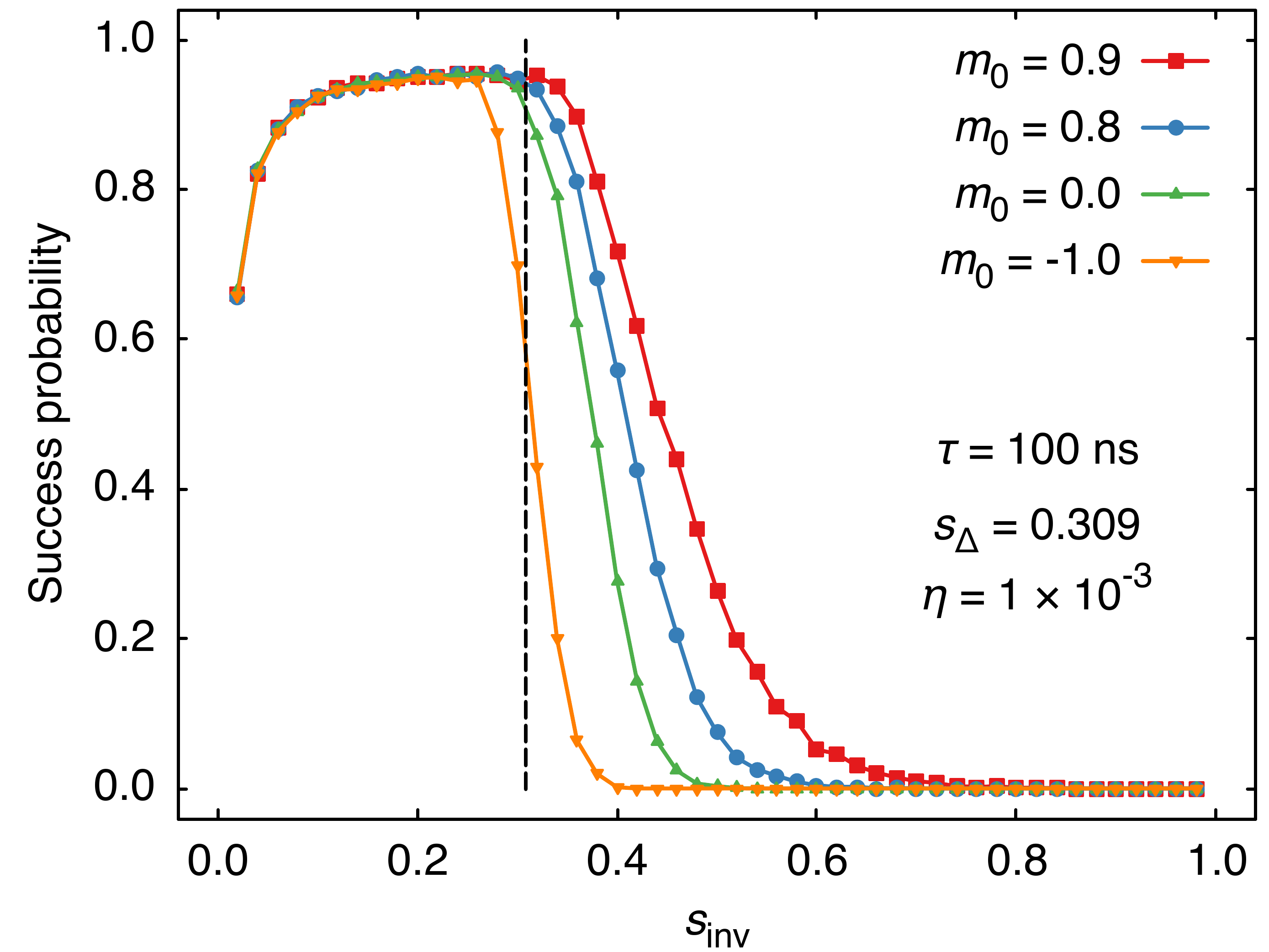}\label{fig:dissipative-tf-100}}
	\subfigure[]{\includegraphics[width = \columnwidth]{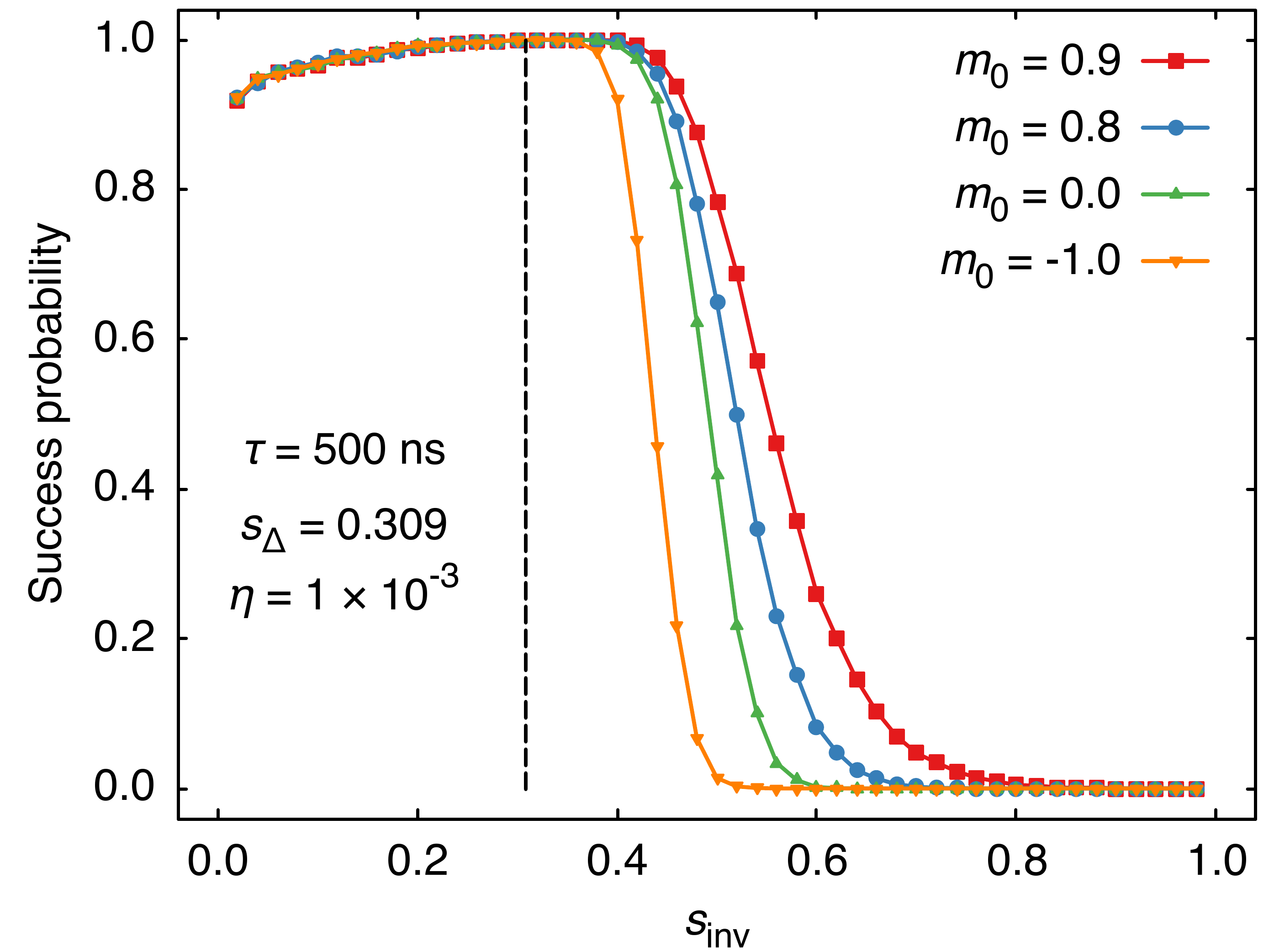}\label{fig:dissipative-tf-500}}
	\caption{Success probability in reverse annealing as a function of the inversion point $ \sinv $, for several values of the magnetization of the initial state. The $ p $-spin system is coupled to a collective dephasing bosonic environment as in Eq.~\eqref{eq:system-bath-collective}, and the coupling strength is $ \eta = \num{1e-3} $. The dashed vertical line denotes the time $ \sgap $ of the avoided crossing between the ground state and the first excited state. In (a) the annealing time is $ \tf = \SI{100}{\nano\second} $, in (b) $ \tf = \SI{500}{\nano\second} $. We sampled the interval $ \sinv \in \rngopen{0}{1} $ using a step $ \Delta s = 0.02 $. All other parameters are given in the main text.}
	\label{fig:dissipative}
\end{figure*}

\section{Open system dynamics subject to dephasing-induced relaxation}\label{sec:dissipative}
	
Physical quantum processors always interact with the surrounding environment, which induces decoherence and thermal excitation/relaxation, which in turn impacts the performance of quantum annealing~\cite{childs:robustness,amin_decoherence_2009,albash:decoherence}. We 
assume weak coupling between the qubit system and the environment. 
It can then be shown that the reduced system density matrix evolves according to a quantum master equation in time-dependent Lindblad form~\cite{zanardi:master-equations}, known as the adiabatic master equation:
\begin{equation}\label{eq:lindblad}
	\frac{d\rho(t)}{dt} = \iu \bigl[\rho(t), \ham(t) + \ham\ped{LS}(t)\bigr] + \diss\bigl[\rho(t)\bigr].
\end{equation} 
In Eq.~\eqref{eq:lindblad}, $ \ham\ped{LS}(t) $ is a Lamb shift term and $ \diss $ is the dissipator superoperator, which makes the dynamics non-unitary and irreversible. They are expressed in terms of Lindblad operators, inducing dephasing or quantum jumps (pumps and decays) between pairs of adiabatic energy eigenstates~\cite{yip:mcwf}. These operators are determined by the instantaneous eigenbasis of the system Hamiltonian $H(t)$ and the system-bath interaction Hamiltonian $ \hamsysbath $. In general, this coupling Hamiltonian involves local operators
that break the spin symmetry of the $ p $-spin model, as each qubit is then coupled to its own bath. We study both this independent decoherence model and the collective decoherence model, wherein  
all the qubits are coupled to a collective bath with the same coupling energy $ g $, in order to preserve the spin symmetry. More specifically, we first consider collective dephasing, for which the system-bath coupling Hamiltonian is
\begin{equation}\label{eq:system-bath-collective}
	\hamsysbath\api{col} = g S_z 
	\otimes B ,
\end{equation}
where $B$ is a bath operator [e.g., $B = \sum_k (a_k + a_k^\dagger)$ for an oscillator bath with annihilation operators $a_k$ for the $k$th bosonic mode].
The Lindblad operators are represented in the instantaneous energy eigenbasis of $H(t)$ as~\cite{zanardi:master-equations}:
\begin{equation}\label{eq:Lab(t)}
	L_{ab}(t) = \braket{E_a(t)| S_z |E_b(t)} \ket{E_a(t)}\bra{E_b(t)}.
\end{equation}
This represents collective dephasing in the energy eigenbasis, wherein the dephasing process randomizes the relative phase between eigenstates of the system Hamiltonian. Thus, this model does not support phase coherence between energy eigenstates.\footnote{It is worth pointing out a caveat. Namely, the collective dephasing model in general supports decoherence free subspaces (DFSs), \ie, subspaces that evolve unitarily despite the coupling to the bath~\cite{Zanardi:97c,Lidar:1998fk}. For instance, the $ S = 0 $ subspace (for even $ \nspin $) is a DFS of the $ p $-spin model. However, the $p$-spin model is unsuitable for performing quantum annealing inside a DFS, since its Hamiltonian consists of operators that preserve the DFS, so that no dynamics would take place if we were to try to encode a computation using states inside the DFS. Instead, to obtain meaningful dynamics (performing a computation) subject to the collective dephasing model, we would need to add Heisenberg exchange terms to the system Hamiltonian~\cite{Kempe:2001uq}.} As a consequence, thermal relaxation tends to equilibrate the system towards its Gibbs state, with a characteristic timescale set by the inverse of the bath spectral density at the gap frequency~\cite{albash:decoherence}.
	
If instead the qubit system is coupled to independent, identical baths, the system-bath coupling operator becomes
\begin{equation}\label{eq:system-bath-independent}
	\hamsysbath\api{ind} = g \sum_{i} \sigma_i^z \otimes B_i, 
\end{equation}
where, e.g., in the bosonic case $B_i = \sum_k (a_{k, i} + a_{k, i}^\dagger)$. Thermal relaxation effects occur here similarly to the collective dephasing case.
However, simulations in this case are more demanding due to the fact that the spin symmetry is broken and that we have $\nspin$ times as many Lindblad operators, \ie,
\begin{equation}
	L_{ab, i}(t) = \braket{E_a(t)| \sigma_i^z |E_b(t)} \ket{E_a(t)}\bra{E_b(t)}.
\end{equation}
Therefore, in this case we will only investigate reverse annealing starting from the first excited state in the symmetric subspace with maximum spin, \ie, $ \ket{w = 1} $, for $ \nspin \in \{3,\dots,8\}$. Moreover, for the particular cases of $ \nspin = 7 $ and $ \nspin = 8 $, we truncate our system to the lowest $ 29 $ and $ 37 $ eigenstates, respectively, to speed up the numerics. This choice is made since the first three levels of the maximum spin subspace at $ s = 1 $ are spanned by $ \sum_{i=0}^2 \binom{7}{i} = 29 $ (for $ \nspin = 7 $) and $ \sum_{i=0}^{2} \binom{8}{i} = 37 $ (for $ \nspin = 8 $) energy eigenstates. We confirm that this is a good approximation by checking that the total population among these levels is close to $1$ during the reverse annealing when additional levels are included in the simulation.
	
\begin{figure}[t]
	\centering
	\includegraphics[width = \linewidth]{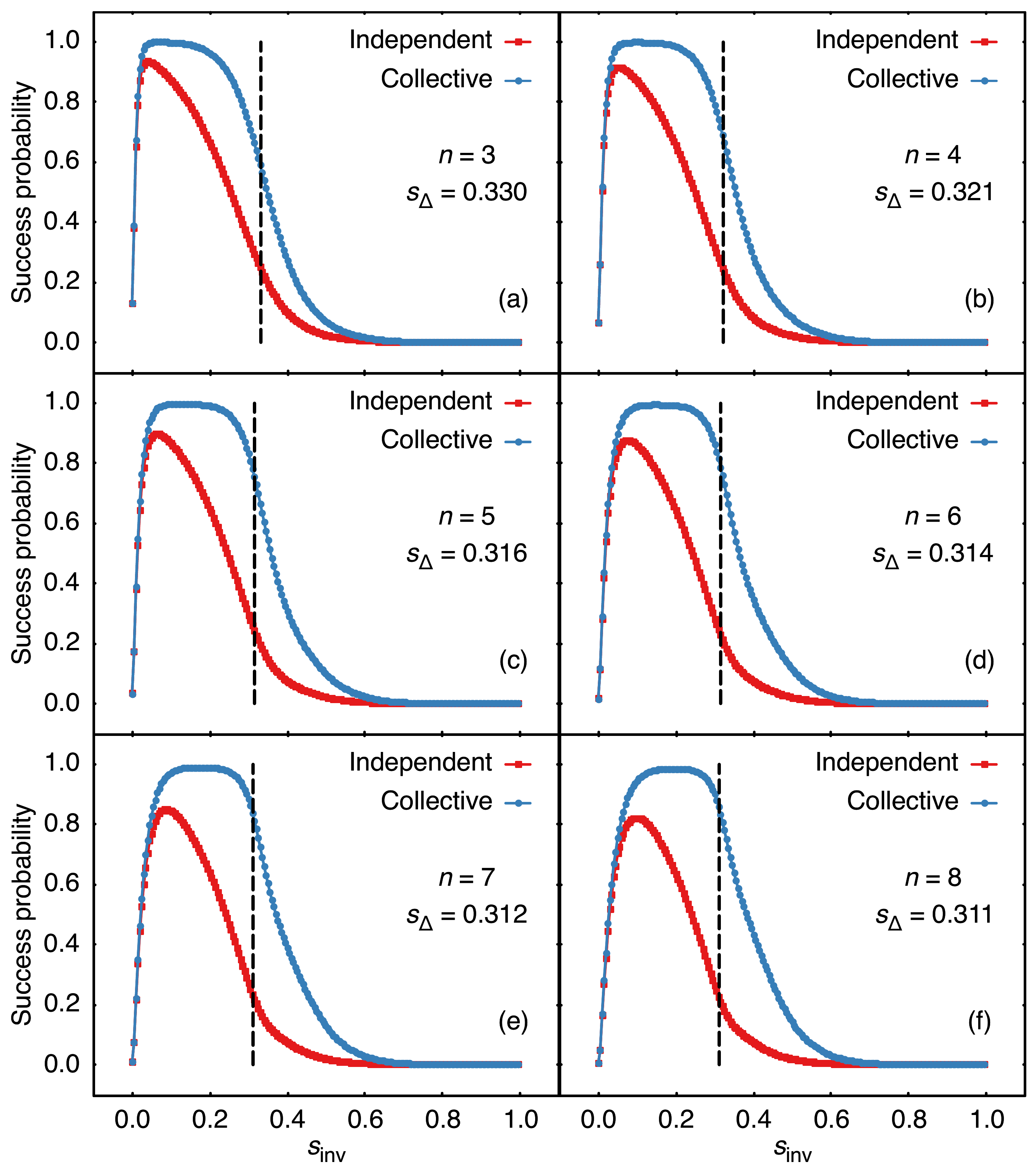}
	\caption{Success probability in reverse annealing as a function of the inversion point $ \sinv $, for $\nspin \in \{3,\dots,8\}$. The initial state is the first excited state of the maximum spin subspace ($ m_0 = 1 - 2/\nspin $). The dashed vertical line denotes the time $ \sgap $ of the avoided crossing between the ground state and the first excited state. The annealing time is $ \tf = \SI{100}{\nano\second} $. We sampled the interval $ \sinv \in (0, 1) $ using a step size of $ \Delta s = 0.005 $.}
	\label{fig:withoutpausing}
\end{figure}

\begin{figure}[t]
	\centering
	\includegraphics[width=\linewidth]{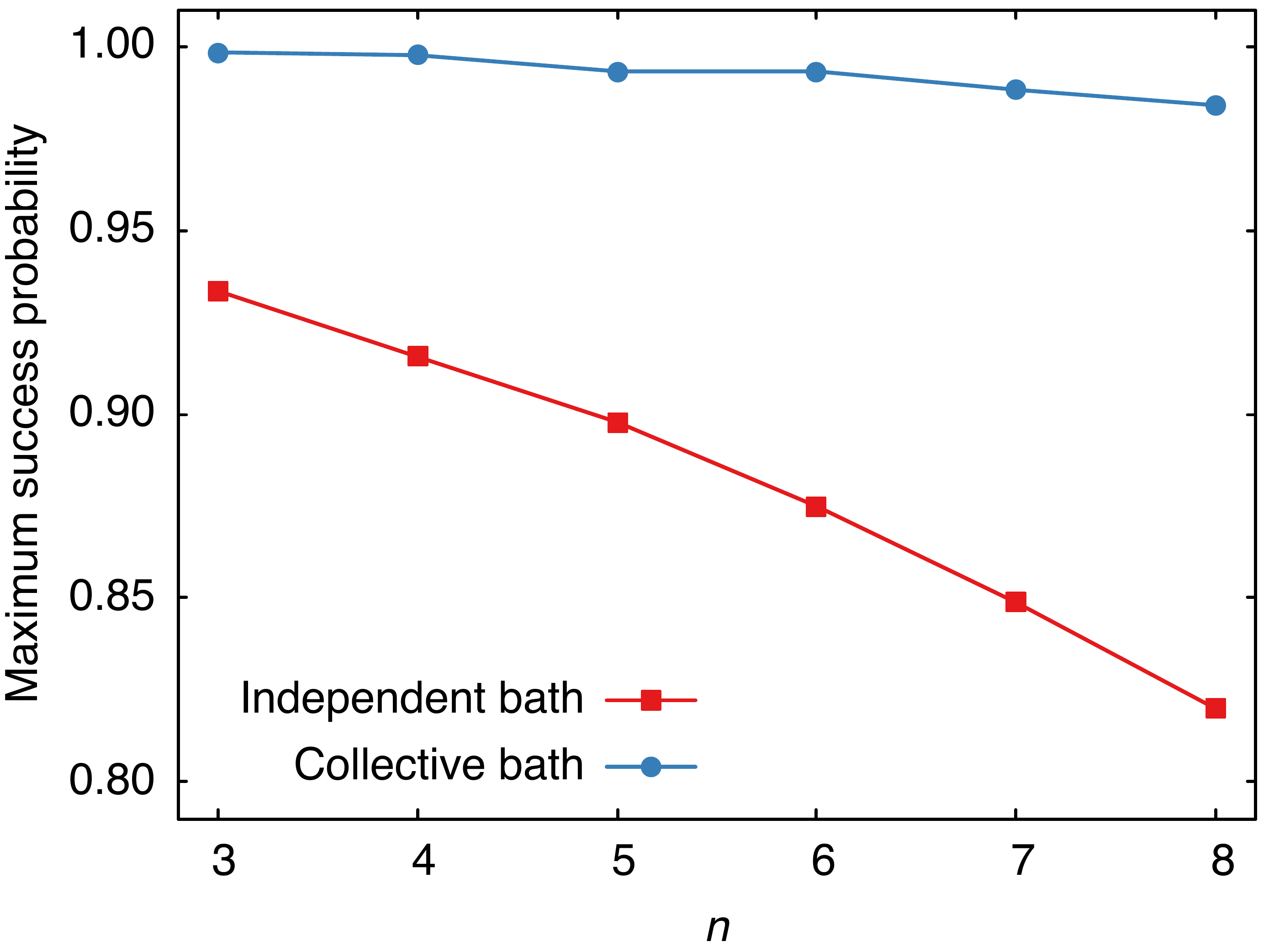}
	\caption{Maximum success probability achievable with reverse annealing as a function of the number of qubits $ \nspin $, using the collective and the independent dephasing models of Eqs.~\eqref{eq:system-bath-collective} and~\eqref{eq:system-bath-independent}, respectively. The annealing time is $ \tf = \SI{100}{\nano\second} $.}
	\label{fig:fiddependenceonn}
\end{figure}

\begin{figure*}[tb]
	\subfigure[]{\includegraphics[width = \columnwidth]{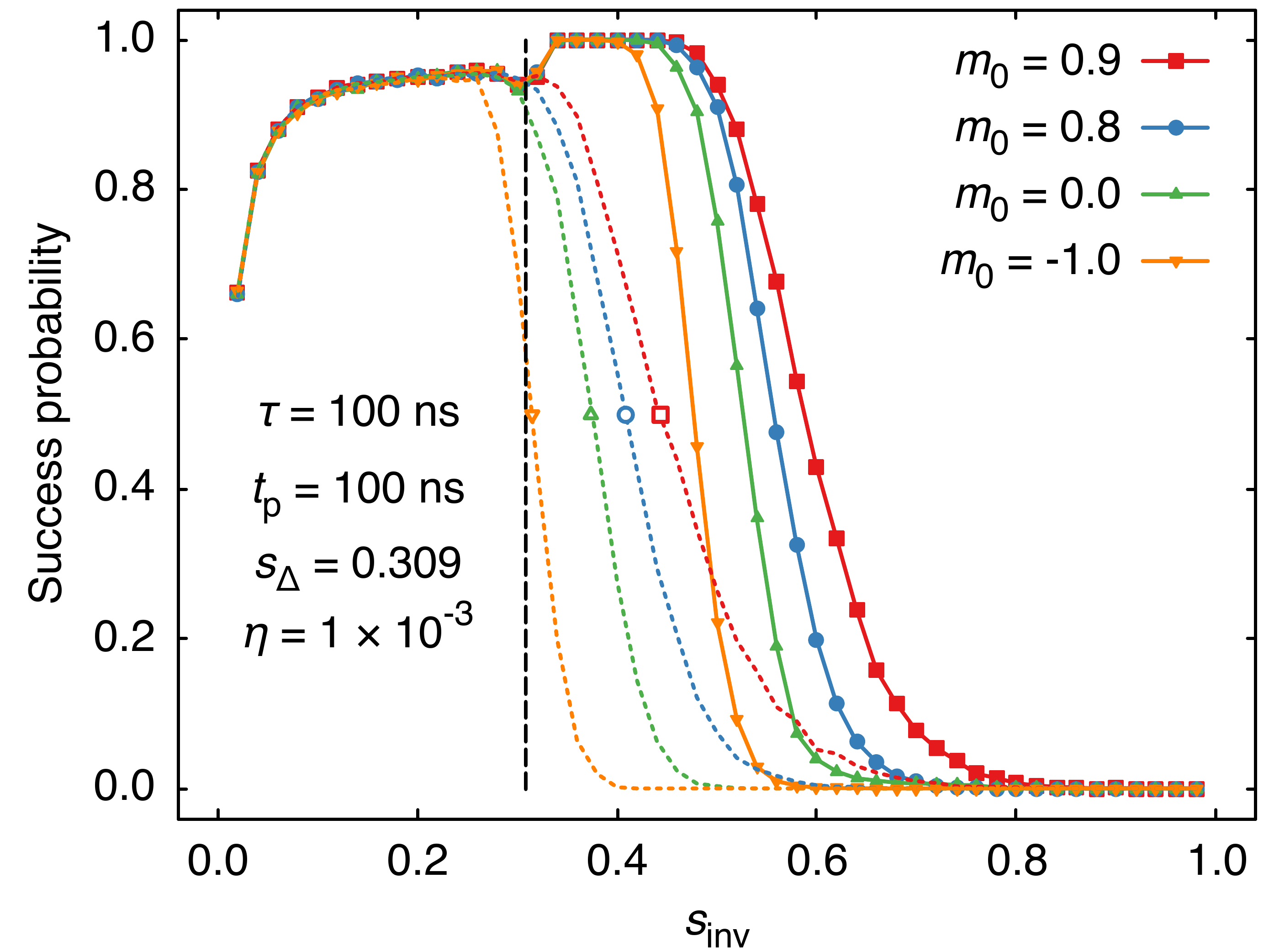}	\label{fig:dissipative-tf-100-pause-100}}
	\subfigure[]{\includegraphics[width = \columnwidth]{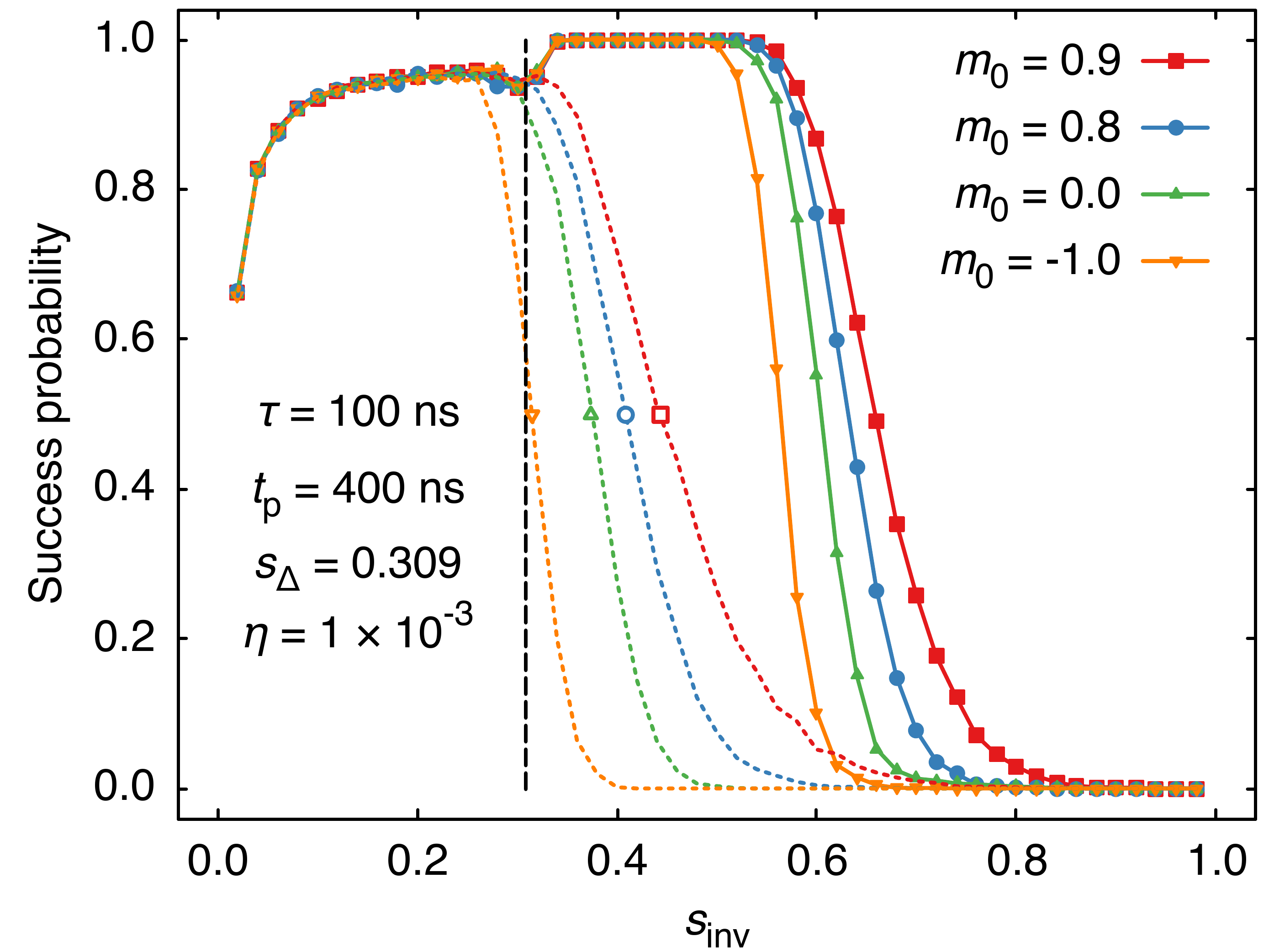}\label{fig:dissipative-tf-100-pause-400}}
	\caption{Success probability in paused reverse annealing for the collective dephasing model, as a function of the inversion point $ \sinv $, for several values of the magnetization of the initial state. The coupling strength is $ \eta = \num{1e-3} $. The annealing time is $ \tf = \SI{100}{\nano\second} $. In (a) a pause of duration $ \lpause = \SI{100}{\nano\second} $ is inserted at the inversion point, while in (b) $ \lpause = \SI{400}{\nano\second} $. All other parameters are given in the main text. Dotted lines with empty symbols refer to open system reverse annealing of time $ \tf = \SI{100}{\nano\second} $ and no pauses. The dashed vertical line denotes $ \sinv = \sgap $. The interval $ \sinv \in \rngopen{0}{1} $ is sampled using a step size $ \Delta s = 0.02 $.}
	\label{fig:dissipative-tf-100-pause}
\end{figure*}		

\begin{figure}[tb]
	\includegraphics[width = \linewidth]{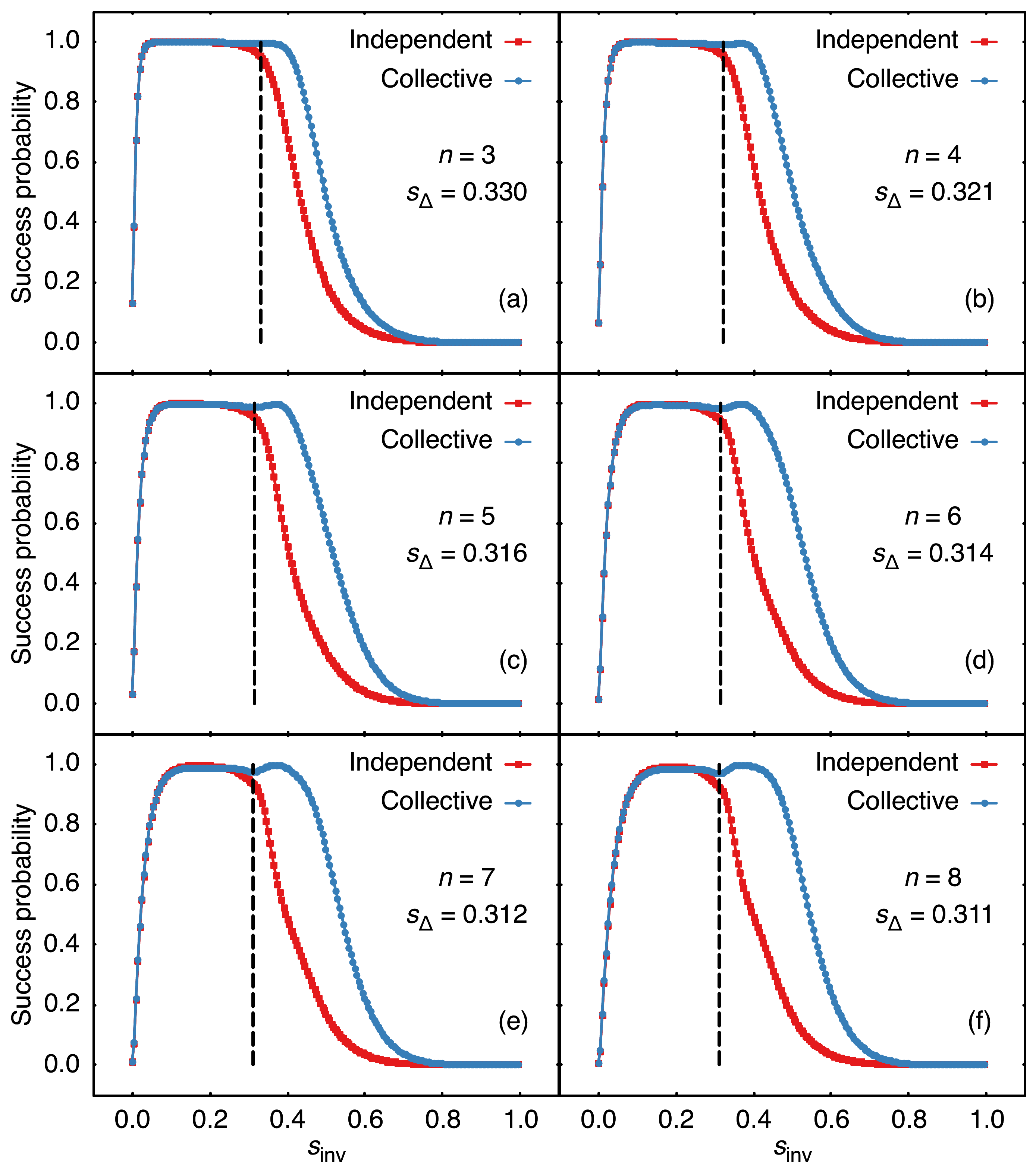}
	\caption{Comparison of the success probability in reverse annealing for the collective and independent dephasing models, as a function of the inversion point $ \sinv $, for $\nspin \in\{3,\dots,8\}$. The initial state is the first excited state of the maximum spin subspace ($m_0 = 1 - 2/\nspin$). The dashed vertical line denotes the time $\sgap$ of the avoided crossing between the ground state and the first excited state. The annealing time is $\tf = \SI{100}{\nano\second}$, and a pause of duration $ \lpause = \SI{100}{\nano\second}$ is inserted at the inversion point.}
	\label{fig:withpausing}
\end{figure}

The adiabatic master equation in Eq.~\eqref{eq:lindblad} is unraveled using a time-dependent Monte Carlo wavefunction (MCWF) approach~\cite{yip:mcwf}. The advantage of MCWF is that it allows to work with wavefunctions rather than density matrices, thus saving quadratically in the dimension of the objects we need to store for numerical calculations. The tradeoff is that to recover the statistical properties of the density operator we need to average over a large number $ K $ of independent trajectories. For the collective system-bath coupling of Eq.~\eqref{eq:system-bath-collective}, the time evolution operator of each trajectory is generated by the effective non-Hermitian Hamiltonian
\begin{align}\label{eq:non-hermitian-hamiltonian}
	\ham\ped{eff}(t) &= \ham(t) + \ham\ped{LS}(t) - \frac{\iu}{2} \sum_{a \ne b} \gamma_{ab} L_{ab}^\dagger(t) L_{ab}(t) \notag \\ & \quad-\gamma_0 \frac{\iu}{2} \sum_{ab}  L_{aa}^\dagger(t) L_{bb}(t),
\end{align}
where $ \gamma_{ab} $ and $ \gamma_0 $ are the rates for jumps and dephasing, respectively. They are related to the temperature and to the spectral density of the bosonic bath,
\begin{equation}\label{eq:spectral-density}
	J(\omega) = g^2 \sum_k \delta(\omega -  \omega_k) = 2\uppi \eta \omega \eu^{-\omega / \omegac},
\end{equation}
where $ \omega_k $ are the bath eigenfrequencies, $ \omegac $ is a high-frequency cutoff and $ \eta $ is the dimensionless coupling strength. We fix $ \omegac = \SI{1}{\tera\hertz} $ and $ \eta = \num{1e-3} $. The working temperature is chosen to be $ T = \SI{12.1}{\milli\kelvin} = \SI{1.57}{\giga\hertz} $, as in experimental quantum annealing systems~\cite{dwave-site}. Eq.~\eqref{eq:non-hermitian-hamiltonian} is easily extended to the independent dephasing case by including a summation over $ i $ for the Lindblad operators and in the Lamb shift term.
	
The time evolution operator generated by the non-Hermitian Hamiltonian of Eq.~\eqref{eq:non-hermitian-hamiltonian} is not unitary. Therefore, the norm of the wavefunction decays in time. Whenever the squared norm decreases below a randomly extracted threshold $ r \in \rng{0}{1} $, a quantum jump occurs, projecting the wavefunction on one of the eigenstates of the Hamiltonian $ \ham(t) $.
The adiabatic master equation dynamics is found after averaging over all stochastic trajectories. In this work, we fix $ K = 5000 $ trajectories and consequently find relative Monte Carlo errors $ \delta\ev{O} / \ev{O} $ of the order of $ \SI{1.5}{\percent} $ over all observables $ O $.
	
We repeat the simulations we reported in Sec.~\ref{sec:unitary} for $ \nspin = 20 $, $ p = 3 $ and $ \tf = \SI{100}{\nano\second} $, but now include the role of the environment. We consider both the collective and independent dephasing models. 
	
In Fig.~\ref{fig:dissipative-tf-100}, we show the success probability as a function of the inversion time $ \sinv $, for the four initial magnetizations $ m_0 = \text{\numlist{0.9;0.8;0;-1}} $. Monte Carlo errors are of the order of the point size in all cases and are invisible. 
	
As in the unitary case, if the inversion occurs too early (\ie, for $ \tinv \ll \tgap $, or, equivalently, $ \sinv \gg \sgap $), the reverse annealing protocol fails to find the ferromagnetic ground state. In fact, thermal excitations are suppressed, as well as Landau-Zener transitions, due to the large level spacing, compared with the temperature and the inverse of the annealing time. For $ \tinv \approx \tgap $ ($ \sinv \approx \sgap $), however, the scenario is drastically different from the unitary case of Fig.~\ref{fig:unitary-tf-100}.

The first difference is that the success probability can be nonzero even if the inversion occurs for $ \sinv \gtrsim \sgap $, especially for $ m_0 = 0.9 $, where the tail of the curve extends to $ \sinv \approx 0.75 $. When the instantaneous gap is of the same order of magnitude as the temperature, thermal processes influence reverse annealing even before crossing the minimal gap. Second, for all $ m_0 $ we observe a sudden increase in the success probability around $ \sinv \approx \sgap $, that eventually brings all curves to an almost flat region at $ \sinv < \sgap $, where the success probability reaches the large value $ \pgs \approx 0.957 $. The value of the maximum success probability at the plateau is $ m_0 $-independent within Monte Carlo errors. The time at which the success probability starts to increase with respect to the baseline depends on $ m_0 $. Moreover, the flat region is wider for larger $ m_0 $, although it has a finite width for all $ m_0 $.
	
These results show that even trial solutions far in Hamming distance from the ferromagnetic ground state can result in a large success probability at the end of a reverse anneal. Moreover, the time window in which inverting the annealing favors the ferromagnetic ordering is relatively large. 
	
We also studied a longer annealing time, $ \tf = \SI{500}{\nano\second} $, as shown in Fig.~\ref{fig:dissipative-tf-500}.
Here, we note that the onset of the success probability plateau shifts towards longer values of $ \sinv $, compared with the $ \tf = \SI{100}{\nano\second} $ case. Therefore, the plateau is wider, and the maximum success probability at the plateau is $ \pgs \approx 1 $ within Monte Carlo errors for all $ m_0 $ we considered. This is in contrast with the unitary case of Fig.~\ref{fig:unitary-scaling}, where increasing the annealing time had detrimental effects on the algorithm. This evidence supports the notion that the success probability enhancement is due to thermal effects, rather than due to purely unitary quantum dynamics~\cite{passarelli:pspin}. Moreover, the adiabatic theorem for open quantum system guarantees convergence to the steady state of the superoperator generator of the dynamics in the large $\tf$ limit~\cite{Venuti:2015kq,Venuti:2018aa}. This too helps to explain our observations: the steady state of the Davies-Lindblad generator of the open system dynamics we considered here is the Gibbs distribution of the final Hamiltonian, which at sufficiently low temperature relative to the gap is the ferromagnetic ground state. Recall that in our case $ \mingap \approx \SI{2.45}{\giga\hertz} $ (at $ \sgap \approx 0.309 $) and $ T = \SI{1.57}{\giga\hertz} $.

We also compare the collective and independent dephasing models of Eqs.~\eqref{eq:system-bath-collective} and~\eqref{eq:system-bath-independent}.
Fig.~\ref{fig:withoutpausing} shows the simulation results for the two models using the adiabatic master equation of Eq.~\eqref{eq:lindblad} for $ \nspin \in \{3,\dots,8\} $. As shown in the figure, simulations using the collective dephasing model have larger success probabilities for almost every $ \sinv$. This is because, in the independent dephasing model, other states not in the subspace of maximum spin become accessible by thermal excitation or diabatic transition during the reverse anneal. For all of the system sizes we simulated, we had to reverse anneal to a smaller inversion point $ \sinv $ for the independent dephasing model to achieve the same success probability as the collective dephasing model. Moreover, the maximum success probability achievable is always smaller for the independent dephasing model. The success probabilities from both models, however, are very similar as $ \sinv \to 0 $, \ie, in the quench limit of the direct part of the evolution.

Figure~\ref{fig:fiddependenceonn} shows how the maximum success probability (over $ \sinv $) of both bath models depends on the number of qubits. As $ \nspin $ increases, the maximum success probability of the independent dephasing model decreases more rapidly than that of the collective dephasing model. While we can infer that if we modeled independent dephasing for $ \nspin = 20 $ we would not observe as large success probabilities as in Fig.~\ref{fig:dissipative}, we stress that  reverse annealing in the independent dephasing model still yields a significantly larger success probability  (for the same $\nspin$ values) than the unitary dynamics case described in Section~\ref{sec:unitary}.

\section{Open system dynamics with a pause}\label{sec:pausing}

Quantum annealing in the presence of a low temperature bath can benefit from pauses inserted at certain times during the dynamics~\cite{marshall, passarelli:pausing}. During a pause, $ s(t) = \text{constant} $, and the system evolves with a time independent Hamiltonian, subject to dephasing. When a pause is inserted some time after $ \sgap $, the environment favors a redistribution of the repopulation according to the Gibbs state at the pause point; at sufficiently low temperature (relative to the gap at this point), this can result in a repopulation of the instantaneous ground state. In this section, we show that pauses at the inversion point can further improve the performance of reverse annealing of the $ p $-spin model.

We repeat the simulations for $ \nspin = 20 $, $ p = 3 $ and $ \tf = \SI{100}{\nano\second} $, using the collective dephasing model. A pause of duration $ \lpause = \tf $ is inserted at $ t = \tinv $, so that the total annealing time, including the pause, is $ \tf' = \tf + \lpause = \SI{200}{\nano\second} $. 
	
In Fig.~\ref{fig:dissipative-tf-100-pause-100}, we report the success probability as a function of the inversion point, for starting magnetizations $ m_0 = \text{\numlist{0.9;0.8;0;-1}} $. We compare the paused case with the unpaused case, for which $ \tf = \SI{100}{\nano\second} $. 
As can be seen in the figure, if the dynamics is reversed too early ($ \sinv \gg \sgap $), the success probability at the end of the anneal vanishes. The level spacing is large compared with the temperature. The relaxation rate is small and the pause is too short to have impact on the dynamics. 
	
However, the presence of a pause significantly changes the outcome of the annealing around $ \sinv \approx \sgap $.  In fact, when a pause is inserted at $ \sinv \gtrsim \sgap $, the success probability reaches $ \pgs \approx 1 $ for a wide range of inversion points and for all $ m_0 $, within Monte Carlo errors. Here, the ground state is completely repopulated by thermal relaxation. This is in contrast with conventional quantum annealing, where the success probability exhibits a peak as a function of the pausing time, when the pause is inserted about $ \SI{20}{\percent} $ later than $ \sgap $, and then rapidly returns to its baseline value~\cite{marshall, passarelli:pausing}.
In contrast, for $ \sinv < \sgap $, the effect of the pause is negligible; the solid (with pause) and dotted (no pause) lines in Fig.~\ref{fig:dissipative-tf-100-pause-100} overlap in this region.

We repeated our analysis for a pause duration $ \lpause = \SI{400}{\nano\second} $, with total annealing time $ \tf' = \SI{500}{\nano\second} $. As shown in Fig.~\ref{fig:dissipative-tf-100-pause-400}, the longer pause duration affects the results only marginally. Comparing with Fig.~\ref{fig:dissipative-tf-100-pause-100}, we note that the qualitative behavior of the curves is the same in the two cases. The pause duration affects mostly the region $ \sinv \gtrsim \sgap $. A longer pause enhances thermal relaxation, thus the success probability starts to increase from its baseline earlier than for shorter $ \lpause $. This results in a wider plateau where the success probability is large, compared with Fig.~\ref{fig:dissipative-tf-100-pause-100}.
	
Finally, we compare the collective and independent dephasing models while including pausing,
starting from the first excited state of the maximal spin sector.
The results are shown in Fig.~\ref{fig:withpausing}, for a pause of duration $ \lpause = \SI{100}{\nano\second} $ inserted at the inversion point. The collective dephasing model continues to exhibit higher success probabilities than the independent dephasing model, as in the case discussed in the previous section, but the results of the two models coincide when $ \sinv < \sgap $. 
Thus, relaxation to the ground state during the pause improves performance for both dephasing models. 
Note that, as $ \nspin $ increases, the maximum success probability of the collective dephasing model is achieved at $ \sinv > \sgap $, while it is achieved at $ \sinv < \sgap $ in the independent dephasing model. This is in agreement with the $ \nspin = 20 $ result shown in Fig.~\ref{fig:dissipative-tf-100}.

\section{Conclusions}
\label{sec:conclusions}

Earlier work revealed an intriguing tension between experimental results demonstrating a substantial enhancement in success probabilities for random spin glass instances under reverse annealing compared to standard (forward) annealing~\cite{marshall}, and theoretical results finding that reverse annealing adversely affects performance for the $p$-spin model, in a closed system setting~\cite{nishimori:reverse-pspin-2}. In this work we resolved this tension by performing a numerical study of reverse annealing of the $p$-spin model in an open system setting, where we included dephasing in the instantaneous energy eigenbasis. We found that the associated thermal relaxation results in significant increase in the success probabilities, as long as the inversion point of the reverse annealing protocol is chosen to be close to the avoided crossing point, or before it. Pausing at the inversion point further improves performance. 

Since closed-system, unitary dynamics predicts that reverse annealing fails, yet its open system analogue succeeds, it follows that thermal relaxation is the mechanism responsible for the success. Reverse annealing is thus an example of a family of protocols that strictly benefit from thermal effects~\cite{verstraete2009quantum,Venuti:2017aa}.
It may be worth noting that quantum effects are likely to play an important role in thermal relaxation because the success probability is very small for $s_{\rm inv}$ close to $1$, i.e. when the Hamiltonian stays classical during the anneal, even with a pause.
Whether this can lead to any quantum speedups is an interesting problem worthy of future investigations.

\acknowledgments
GP and PL acknowledge the CINECA Award under the ISCRA initiative (project IscraC\_QA-MCWF) for the availability of high-performance computing resources and support. They also acknowledge fruitful discussions with Prof.~Vittorio Cataudella and Prof.~Rosario Fazio. 

The research of KY, DL, and HN is based upon work (partially) supported by the Office of
the Director of National Intelligence (ODNI), Intelligence Advanced
Research Projects Activity (IARPA), via the U.S. Army Research Office
contract W911NF-17-C-0050. The views and conclusions contained herein are
those of the authors and should not be interpreted as necessarily
representing the official policies or endorsements, either expressed or
implied, of the ODNI, IARPA, or the U.S. Government. The U.S. Government
is authorized to reproduce and distribute reprints for Governmental
purposes notwithstanding any copyright annotation thereon. Computation for some of the work described in this paper was supported by the University of Southern California Center for High-Performance Computing and Communications (hpcc.usc.edu).
	
%

\end{document}